# Using consumer behavior data to reduce energy consumption in smart homes

Applying machine learning to save energy without lowering comfort of inhabitants

Daniel Schweizer, Michael Zehnder, Holger Wache, Hans-Friedrich Witschel
Institute of Business Information Systems
University of Applied Sciences and Arts Northwestern Switzerland
FHNW - Olten, Switzerland
mzehnder@gmx.net
{holger.wache;hansfriedrich.witschel}@fhnw.ch

Danilo Zanatta, Miguel Rodriguez
Research and Development
digitalSTROM AG
Zurich, Switzerland
{danilo.zanatta;miguel.rodriguez}@digitalstrom.com

*Abstract*— **This paper discusses how usage patterns and preferences of inhabitants can be learned efficiently to allow smart homes to autonomously achieve energy savings. We propose a frequent sequential pattern mining algorithm suitable for real-life smart home event data. The performance of the proposed algorithm is compared to existing algorithms regarding completeness/correctness of the results, run times as well as memory consumption and elaborates on the shortcomings of the different solutions.**

**We also propose a recommender system based on the developed algorithm. This recommender provides recommendations to the users to reduce their energy consumption. The recommender system was deployed to a set of test homes. The test participants rated the impact of the recommendations on their comfort. We used this feedback to adjust the system parameters and make it more accurate during a second test phase. The historical dataset provided by digitalSTROM contained 33 homes with 3521 devices and over 4 million events. The system produced 160 recommendations on the first phase and 120 on the second phase. The ratio of useful recommendations was close to 10%.**

## I. Introduction

King defines a smart home as a "dwelling incorporating a communications network that connects the key electrical appliances and services, and allows them to be remotely controlled, monitored or accessed" [1]. The definition goes on by stating benefits that a smart home can have regarding energy management like running appliances when energy is cheapest or controlling the air conditioning/heating for maximum efficiency when the house is busy or empty.

Harper acknowledged that "smart house technologies that most people are pleased with are connected with saving energy or money" [2]. Moreover, previous studies carried by digitalSTROM show that energy savings can only be achieved if the inhabitants comfort is taken into consideration [3].

Based on these findings, we believe that a smart home needs to be able to predict future needs of its inhabitants before initiating actions to reduce the energy consumption without decreasing the comfort of its inhabitants. While Wang [4] suggest that manually specified preferences are used, this project wants to analyze if and how such preferences in the form of frequent and periodic patterns can also be gained autonomously through sequential pattern mining. Frequent means a pattern which occurs more often than others, while periodic means a pattern which occurs at a constant interval. Manual specification of such rules is not desirable because of the high degree of interaction and involvement required from the inhabitants.

In this paper we propose an autonomous recommender system to provide energy saving actions to the home inhabitants without reducing their comfort level. In order to achieve this goal, the recommender has to be fed with the event stream from the smart home and to previously mined patterns that reflects the inhabitants' behavior. We propose thus a frequent sequential pattern mining algorithm tailored for mining smart home event data.

The rest of this paper is organized as follows. Section II provides some background about smart homes and pattern mining applied to them. Section III describes the proposed algorithm for mining patterns in smart homes events. Section IV presents the proposed recommender to reduce energy consumption in the smart homes. The results of field tests are presented in Section V and in Section VI conclusions are drawn and future work directions are described.

## II. Background

### A. Pattern mining for smart homes

A suitable algorithm allows a smart home to learn its inhabitants' usage patterns autonomously. The Apriori algorithm was originally designed with the goal of association rule mining in mind, as described by Agrawal & Srikant [5]. In the smart home context, the sequence of events is of crucial importance when analyzing the data generated by the inhabitants. In order to learn from the inhabitants habits, it is relevant if appliance A was turned on before or after appliance

B. However, the Apriori algorithm does not take into account the sequence of events when detecting patterns.

Rashidi and Cook [6] adapted the Apriori algorithm to consider the order of events when mining patterns from smart home data. They look for patterns that are at least two activities long and then iteratively extend the length of the patterns by one until they are unable to find frequent patterns anymore.

A wide range of sequential pattern mining algorithms are implemented in the open-source data mining library SPMF by Fournier-Viger et al. [7]. The implemented generic algorithms include BIDE+ [8] and PrefixSpan [9]. They are in general considered very fast and efficient sequential pattern mining because of their usage of pruning techniques.

Another desired feature that a pattern mining algorithm used for smart homes should support is wildcarding. The idea behind wildcarding patterns is that there are interesting patterns that are very similar, differing only by a few events (e.g. one or two), but are not considered frequent because of this variation. For the sake of clarity, consider a set of patterns that differ by only one event. If a wildcard would replace this event, the frequency of occurrence of such pattern would increase and the wildcarded pattern would become frequent. GapBIDE by Li [10] is an adaption and enhancement of the previously mentioned BIDE+ algorithm, taking wildcarding into account. We use GapBIDE as a representative of sequential pattern mining algorithms. Other algorithms in this class are, for example, MAIL by Xie [11] and PMBC by Wu [12].

B. Smart home technology

digitalSTROM products provide connectivity to electrical devices in the home over the existing power cables. This includes every lamp, light switch, blinds and any plugged in device. This network of devices is connected through a server mounted in the electrical cabinet to a local area network. The result is a network of connected devices, bringing the internet of things (IoT) to each home. digitalSTROM components are based on a high volt IC in a small size module. Each digitalSTROM module can switch, dim, measure electricity and communicate its status. The products are available through Europe with its larger installed base in Germany and Switzerland.

A digitalSTROM system is based on concentrators that reside in the electrical distribution panel, acting as power meters for the individual distribution circuits and communicating with individual nodes installed within a home over differently modulated up and downstream channels [13]. The system includes a Linux server application with a JSON API. Moreover, real-time data from test homes are collected by a logging system, parsed and stored in a database, being available for processing. From this logging system, historical data can easily be obtained and processed.

C. Criteria used for choosing pattern mining algorithms

The choice of the pattern mining algorithms used in this work was influenced by the following criteria:

- The algorithm has to be able to find patterns in data. To be more precise it has to find both frequent sequential patterns and periodic sequential patterns

- It has to be able to find wildcarded patterns and to output where the wildcard is positioned in the pattern

- The algorithm needs to be able to process the continuous stream of data coming from a smart home, i.e. it must be able to process this events in real-time

In general both Apriori-inspired algorithms employing pruning techniques, from the family of deterministic algorithms, as well as genetic algorithms, from the family of heuristic algorithms, are suitable for finding frequent patterns in large datasets. In this work, we consider only deterministic algorithm, since they are able to find patterns in a reasonable amount of time and do not have the disadvantage of getting stuck at local maxima.

D. Training data set

The historical training dataset used contains 33 homes with 3521 devices, which are related to 4,331,443 events and 6829 unique scenes. These events extend over a period between 08/12/2002 and 25/06/2014. There are several types of electrical devices that can be connected and they are grouped by functionality, such as lighting, shades, audio and so on. The events themselves are either in one of the device groups or are of a broadcast or unknown type.

Events have a scene ID and a source ID. The scene ID defines the action executed, like "Turn light in kitchen off", while the source ID can be used to identify if it was a user generated event, a scripting event, e.g. the one generated by a system timer, or a sensor event, e.g. the one generated by a motion detector or temperature sensor.

For example, the following 3 events compose one pattern:

| Time | Event data | | |
|---|---|---|---|
| | Event | Source ID | Scene ID |
| 28.04.2012 13:26:38 | Turn on the light in living room to evening light scene | 377 | 434 |
| 28.04.2012 13:30:39 | Turn on the light in living room to the reading scene | 381 | 424 |
| 28.04.2012 13:41:50 | Turn off light in living room | 381 | 422 |

III. ALGORITHM DESIGN

A. Window Sliding with De-Duplication (WSDD)

A new algorithm for mining patterns from smart home events was designed. The algorithm is based on the idea of so-called open and overlapping patterns [14]. It follows a brute force approach, namely find all possible frequent patterns by sliding a window of a specifiable size over the chronologically ordered events and count their overall occurrences inside the data. The patterns found by this algorithm were encouraging: a two-event pattern with over 9% support was found and supported the assumption that patterns are deducible in event data of smart homes. However, the run times were long, as expected from a brute-force approach.

In order to improve the algorithm, a hash map data structure was used. This hash map uses the key for storing the pattern itself and the support count of this pattern is stored as the value. This means that, when the proposed algorithm has finished mining, the support of a certain pattern is stored as the

value, while the pattern itself is the key. Therefore, we achieve a speed improvement by doing the following:

1. Pattern de-duplication: The enhanced algorithm avoids mining patterns more than once. This gives the algorithm its name: WSDD, Window Sliding with De-Duplication

2. Instead of building possible patterns in a first loop and then mine for these found patterns in a 2nd loop, the counting is done in the same loop. This eliminates the need for re-iterating over the events

As an additional improvement, in analogy to the Apriori algorithm, a minimum support parameter was introduced. This was done because very infrequent patterns are not interesting, e.g. all the patterns that occur only once can never be considered *normal behavior* of the smart home inhabitants. The minimum support parameter is therefore used to post-process the patterns found. This algorithm returns only patterns that have a support greater than this minimum.

The final proposed algorithm is depicted in Fig. 1.

*B. Algorithm comparison*

To compare the WSDD algorithm with the two well-known sequential pattern mining algorithms BIDE+ and PrefixSpan, an interface to the open-source Java framework SPMF by Fournier-Viger et al. [7] was implemented. It became obvious very soon that regarding throughput and run times, the developed WSDD algorithm is very competitive for the available data set. For a minimum support of 0.01, PrefixSpan needs 3 seconds for a smart home with 80000 events, BIDE+ needs 25 seconds while WSDD finishes in less than 1 second.

A reason for the good run times of the WSDD algorithm is that the number of the different patterns in a smart home is relatively small which allows the algorithm to store all the patterns without producing very big hash maps.

Moreover, there are three general advantages of the smart home specific algorithm WSDD over general purpose algorithms such as BIDE+, PrefixSpan and GapBIDE:

- WSDD reports not only frequent but also periodic patterns
- WSDD reports where wildcards are detected in a pattern
- No post-processing is needed for the correct support count

These results were expected given the fact that PrefixSpan, BIDE+ and GapBIDE are generic algorithms, not explicitly developed for frequent sequential pattern mining in smart homes.

We ran benchmarks measuring both the run times as well as the memory consumption. The different algorithms were run multiple times and with various different parameter settings. To get representative results, a selection of five different smart homes with various amounts of events was chosen: the smart home with the fewest events (1,173 events), three smart homes with average amounts of events (22,489 – 42,129 events) and the smart home with most events (156,121 events). The results of an example run can be seen in Fig. 2.

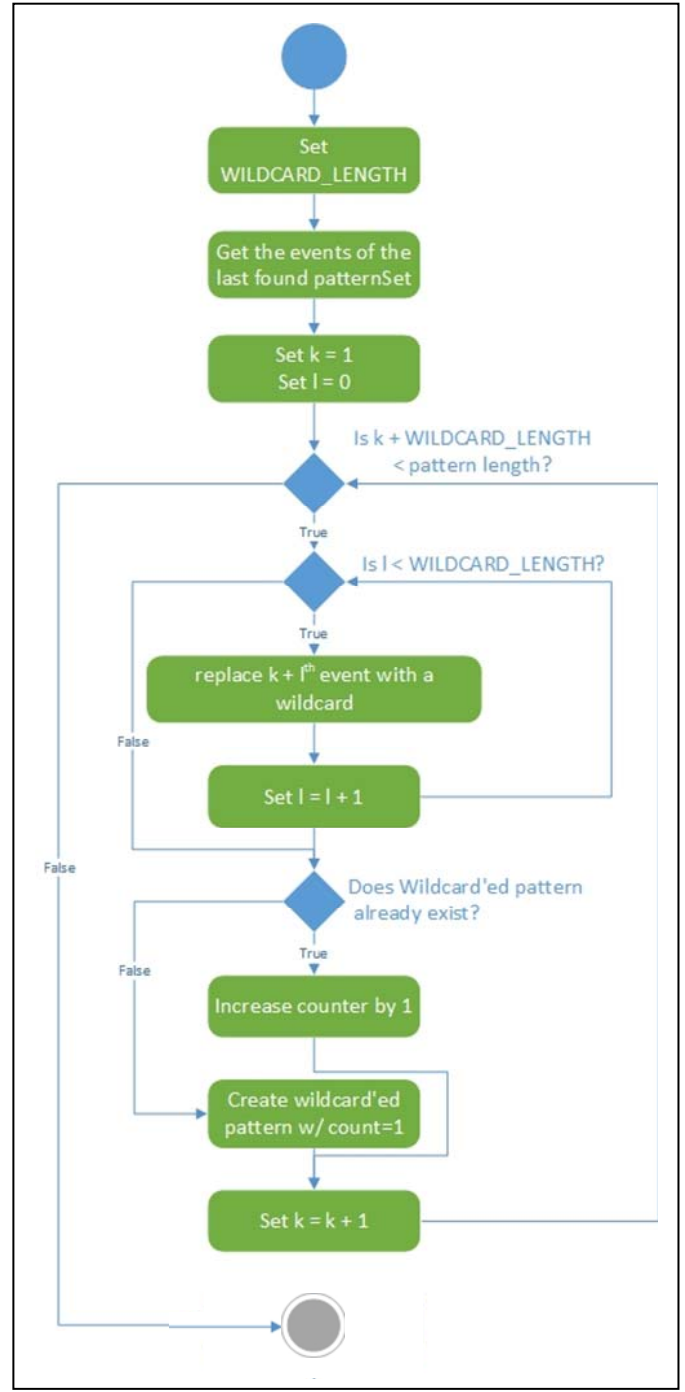

Fig. 1. WSDD algorithm

In addition to the run time benchmarks, the same four algorithms were compared regarding their memory consumption. The memory consumption was measured with the "MemoryLogger", which is provided as part of the SPMF framework and with Performance Monitor of Windows 8. The benchmark settings were the same as for the previous run time experiments.

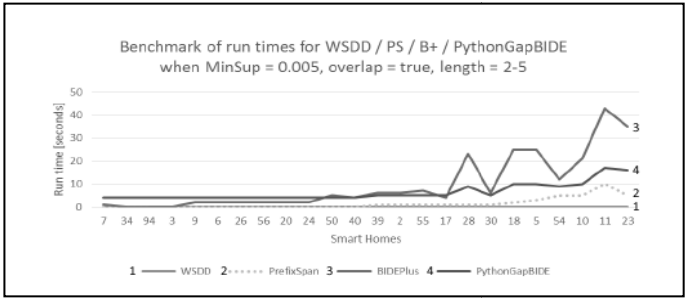

Fig. 2. Benchmark of run times for data mining algorithms

TABLE I. TOP 20 WILDCARDED PATTERNS OF ONE SMART HOME

| Support Count | 1. | 2. | 3. | 4. | 5. |
|---|---|---|---|---|---|
| 8706 | a | b | | | |
| 8495 | b | a | | | |
| 8141 | a | * | a | | |
| 7913 | a | b | a | | |
| 7807 | b | * | b | | |
| 7787 | b | a | b | | |
| 7488 | a | * | a | b | |
| 7335 | a | b | * | b | |
| 7324 | b | a | * | a | |
| 7320 | a | b | a | b | |
| 7169 | b | * | b | a | |
| 7153 | b | a | b | a | |
| 6905 | a | b | a | * | a |
| 6904 | a | * | a | b | a |
| 6806 | b | a | * | a | b |
| 6767 | a | b | * | b | a |
| 6755 | a | b | a | b | a |
| 6692 | b | * | b | a | b |
| 6689 | b | a | b | * | b |
| 6677 | b | a | b | a | b |

The results from memory consumption show that PrefixSpan, GapBIDE and, especially WSDD, are significantly better than BIDE+. The memory consumption of the former three algorithms ranges from a few dozen MB to some hundred MB, depending on the number of events, the minimum support and the pattern length. BIDE+ on the other hand uses up to 1.7 GB.

It can be concluded that, from a memory consumption point of view, WSDD and GapBIDE have the best results, while PrefixSpan uses slightly more memory and BIDE+ uses significantly more memory.

The correlation between the input parameters and the memory consumption can be clearly observed: the more events, the longer the patterns and/or the less the minimum support, the higher the memory consumption.

An additional insight into memory consumption was that the increase of events and the lowering of the minimum support favor WSDD more than GapBIDE, meaning that the increase of memory consumption for GapBIDE is proportionally higher for more events and/or less minimum support than it is for WSDD.

A comparison of wildcarded patterns with non-wildcarded patterns does not show significant improvement. As TABLE I. shows, there are only few instances where wildcarding produces significantly different results from when wildcarding is deactivated. Furthermore the increase of the support count of the wildcarded patterns over the non-wildcarded patterns is mostly rather low. It can therefore be said that wildcarding is not worth considering when mining frequent patterns in smart home event data since the increase in run times and memory consumption is not yielding significantly better results. In the following we consider only non-wildcarded pattern.

IV. DESIGN OF THE RECOMMENDER SYSTEM

Because not all frequent or periodic patterns result in energy savings, we defined some characteristics to identify the relevant behavior patterns.

To ensure that a *relevant pattern* can be used to save energy, it must be composed of two main components:

(1) A relevant pattern must contain at least one action to lower energy usage (actions are a subset of normal events). For a detailed description, see [15]

(2) The pattern must consist of *normal events*, which serve as condition to suggest the action at the right time.

Because a relevant pattern consists of normal events, which represents the condition, and an action, this pattern can be interpreted using an association rule. An association rule is an implication of the form

$$X \rightarrow Y, \text{ where } X, Y \subset I,$$

where X is a sequence of normal events, Y is a single action, and I is the set of all possible events. The association rule above states that when X occurs, Y occurs with certain probability [16].

For this application no motion detectors or other location sensors where used. Only activities like turning on (or off) light, TV etc. are considered.

A. Architecture

We proceeded to build a recommender system using the WSDD algorithm we designed. The architecture of the recommender system developed in this project is shown in Fig. 3 and can be divided in three main parts:

- The storage of the association rules
- The event stream of the current behavior data inside the smart home
- The matching algorithm

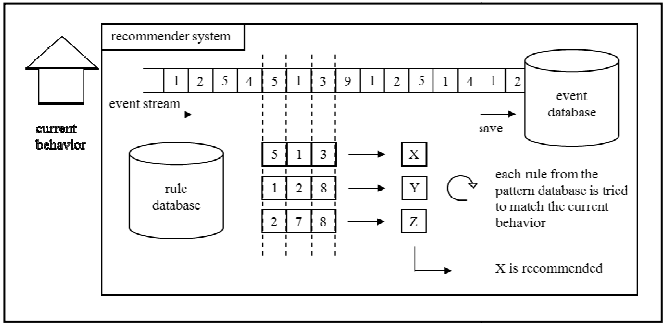

Fig. 3. Recommender system architecture

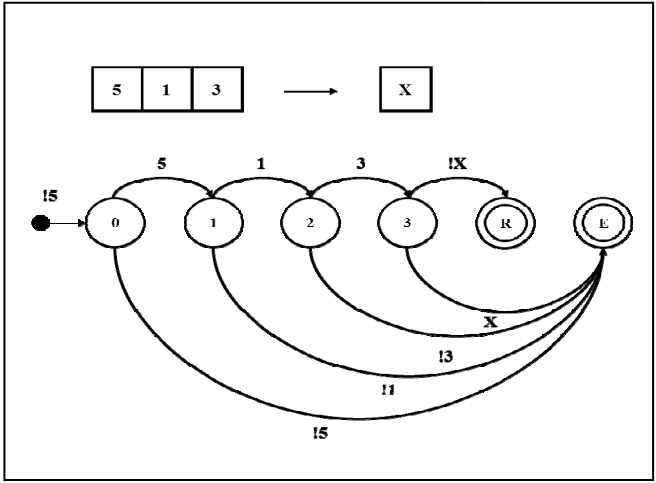

Fig. 4. Example of a matching rule in the algorithm (FSM)

The **rule database** stores the association rules, which were obtained from the mined relevant patterns. The **event stream** contains the real-time events from the smart home, ordered by time of their occurrence.

The **matching algorithm** is the core component of the recommender system. It matches the rules and the event stream. The most common existing rule matching algorithm is RETE by Forgy [17]. We use a deterministic finite state machine (FSM) approach as depicted in Fig. 4, which reflects the order of the events better than RETE. A new instance of the FSM is created for each new event in the stream. If there is no matching in the first attempt, the instance is removed from memory. If the condition did match and the next event is not the action itself, the machine sends a recommendation.

The design of the recommender system allows more than one rule to be matched at the same time. In order to avoid multiple conflicting recommendations, we propose to weight the rules and use the weights as a prioritization criterion.

### B. Implementation

The recommender system was implemented on a cloud based Microsoft® Azure VM (Virtual Machine). The VM was set up with Ubuntu 14.04.

The smart homes event data is parsed from the log files of the digitalSTROM system. The files are uploaded by a script installed on the digitalSTROM infrastructure in the houses and made accessible for this project on a file server. The files are copied every 5 minutes by remote synch to the VM where the recommender system is running. They are parsed by a Python script and stored in a MySQL database.

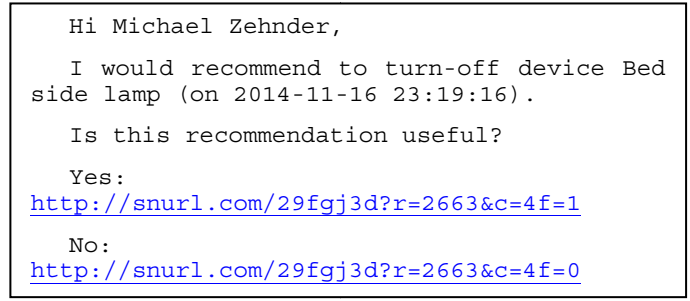

Fig. 5. Recommendation SMS

## V. RESULTS

The evaluation was conducted in 33 households with inhabitants producing real event data. The test households were equipped with the smart home automation system digitalSTROM. The historical event data of the homes was mined by the recommender system in advance. According to the number of relevant patterns found in the houses, they were ranked and the owners of the 15 most promising households where requested to take part in the evaluation. From these, 8 houses agreed to participate in the evaluation of this work including both single- and multi-inhabitant houses. Recommendations were sent per SMS to the mobile devices of the inhabitants. An example of such recommendation is shown in Fig. 5.

We ran the evaluation in two phases, which are described in the following sections.

TABLE II. KEY RESULTS OF EVALUATION

| Parameter | Phase | |
|---|---|---|
| | *1* | *2* |
| # days evaluated | 14 | 34 |
| Recommendations sent | 160 | 120 |
| Answered recommendations | 76 | 55 |
| Voted useful | 7 | 5 |
| Voted not useful | 69 | 50 |
| Ratio useful/answered | 9.21% | 9.10% |
| Number of active rules | 54 | 46 |
| Number of rules that resulted in recommendations | 23 | 17 |
| Number of rules with 10 negative feedbacks | 5 | 3 |

### A. Phase 1

The aim of the first phase was to provide a large basis of data for evaluation and further improvement of the system. The analysis of the data collected during phase 1 should help to improve the recommender system in terms of decreasing the negatively rated recommendations in phase 2, while holding the positives at the highest amount possible. The results for phase 1 are summarized in TABLE II.

After running the evaluation for phase 1 for 2 weeks, the inhabitants were interviewed and their feedback was used to improve the recommender for the second phase. We also did a regression analysis of the results using the *weighted feedback* as the dependent variable and the following prioritization served as explanatory variables:

- The length of the pattern
- The position of the action
- Support of the pattern
- Confidence of the rule

The result of the regression analysis shows that patterns with high confidence and high pattern-length tend to receive better feedback than the other patterns. On the other hand, support and position of the action did not show any significance to describe the feedback a rule. It is worth to notice that support was the major attribute for mining frequent patterns. Fig. 6 shows the regression based on confidence.

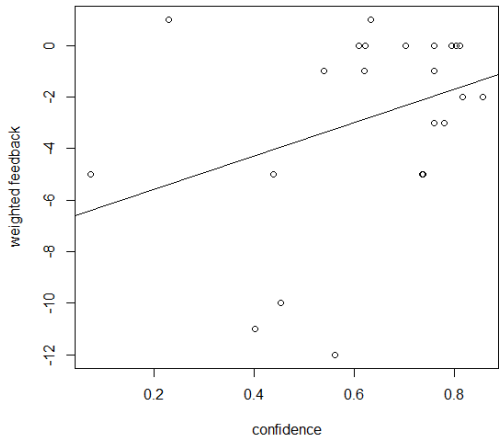

Fig. 6. Scatter plot of confidence with line of best fit

### B. Phase 2

The aim of the second phase of the evaluation was to increase the ratio of answered recommendations and the ratio of useful recommendations compared to the first phase. The analysis of the data collected during phase 1 was used to adapt the system, which should lower negative rated and unanswered recommendations. The prototype was improved in the following points:

- All rules with 10 negative feedbacks in a row during phase 1 were removed from phase two
- To mitigate the problem of the low response rate caused by ambiguous recommendations, we enriched the text with the name of the room where the device is located
- As result of the regression analysis, confidence and pattern-length of each rule where multiplied with their estimate to calculate a coefficient which gives indication about the *usefulness* of a rule. A threshold is defined and 19 rules out of 54 were excluded from the second phase (35 rules remained)

The results from phase 2 are summarized in TABLE II. The results show a similar ratio of useful recommendations as in phase 1, as well as a similar response rate (45.8%). However, a significant improvement can be observed in the number of recommendations sent: 0.44 recommendations/day/home in phase 2 *versus* 1.43 recommendations/day/home in phase 1. Note that, for the same ratio of useful recommendations, a lower number of recommendations per day per home means less *noise* for the user and a better comfort level.

We consider the 10% ratio of useful recommendations a promising good start for a first version of the system which has not been optimized nor has it seen a large amount of usage data.

## VI. CONCLUSION AND FUTURE WORK

In this work, we proposed an algorithm to mine data from smart home events and a recommender system that helps to save energy in smart homes without reducing the comfort of the inhabitants.

The proposed algorithm, called WSSD, outperforms existing algorithms both in run-time as well as in memory usage. Even though WSDD allows the mining of wildcarded patterns, their use does not show any significant improvement over non-wildcarded patterns for smart home events, with the drawback of increased memory usage and runtime.

Based on this algorithm, we proposed a recommender system that generates recommendations for the smart home inhabitants with the aim of reducing power consumption without decreasing their comfort. The results show that such a system works in *real life* and achieved a ratio of useful recommendations of about 10%, while sending 0.44 recommendations/day/home.

Several points of improvement were identified during the evaluation phases of this work. A follow-up research project is already ongoing and will build upon the findings of this work. The following ideas for further research materialized during the design, implementation or evaluation of the recommender system:

- Using confidence and pattern length instead of support or periodicity as criteria for the mining algorithm, resulting in more and better patterns
- The time between two events (or the action) is considered neither by the mining algorithm nor by the recommender system. Using this information will improve the accuracy of the suggestions made by the system
- Other attributes could be introduced to decide if a rule is relevant or not. Such attributes might be:
    - Time of day when the pattern occurs most
    - Weekday when the pattern occurs most
    - Season when the pattern occurs most
- The recommender should learn from the feedback of the inhabitants in order to prioritize the rules, instead

of just excluding a rule after 10 negative feedbacks in a row

- Look into estimating the amount of energy that would be saved by the recommendations
- Test other machine learning algorithms and frameworks such as "Torch and Caffe"